\documentclass[10pt,conference]{IEEEtran}
\usepackage{cite}
\usepackage{amsmath,amssymb,amsfonts}
\usepackage{algorithmic}
\usepackage{graphicx}
\usepackage{textcomp}
\usepackage{xcolor}
\usepackage[hyphens]{url}
\usepackage{booktabs}
\usepackage{pifont}
\usepackage{siunitx}
\usepackage{amsmath}
\usepackage{algorithm2e}
\usepackage{multirow}
\usepackage{fancyhdr}
\usepackage{float}
\usepackage{setspace}
\usepackage{listings}
\usepackage{balance}
\usepackage{wrapfig}
\usepackage{marginnote}
\usepackage{caption}
\usepackage{floatflt}
\definecolor{gfored}{rgb}{0.580, 0.050, 0.211}
\definecolor{ao}{rgb}{0.007, 0.520, 0.867}
\definecolor{yt}{rgb}{0.875, 0.568, 1.000}
\definecolor{moegi}{rgb}{0.357, 0.537, 0.188}
\definecolor{jl}{rgb}{1.0, 0.2, 0.8}
\definecolor{brown(web)}{rgb}{0.65, 0.16, 0.16}
\definecolor{bisque}{rgb}{1.0, 0.89, 0.77}

\newif\ifsqueezefigs
\squeezefigstrue

\ifsqueezefigs
\makeatletter
\g@addto@macro{\normalsize}{%
  \setlength{\abovedisplayskip}{2pt plus 1pt minus 1pt}
  \setlength{\belowdisplayskip}{2pt plus 1pt minus 1pt}
  \setlength{\abovedisplayshortskip}{0pt}
  \setlength{\belowdisplayshortskip}{0pt}
  \setlength{\intextsep}{2pt plus 1pt minus 1pt}
  \setlength{\textfloatsep}{3pt plus 1pt minus 1pt}
  \setlength{\dbltextfloatsep}{3pt plus 1pt minus 1pt}
  \setlength{\skip\footins}{4pt plus 1pt minus 1pt}}
\makeatother
\fi

\newif\ifdraft
\draftfalse

\ifdraft
    \usepackage[colorinlistoftodos,prependcaption,textsize=small]{todonotes}
    \paperwidth=\dimexpr \paperwidth + 4cm\relax
    \oddsidemargin=\dimexpr\oddsidemargin + 2cm\relax
    \evensidemargin=\dimexpr\evensidemargin + 2cm\relax
    \marginparwidth=\dimexpr \marginparwidth + 2cm\relax

    \newcommand{\ominline}[1]{\textcolor{red}{\textbf{[@om: }#1\textbf{]}}}
    \newcommand{\ombox}[1]{\todo[size=\scriptsize, linecolor=red, bordercolor=red, backgroundcolor=white]{\textcolor{red}{\textbf{@om:} #1}}}

    \newcommand{\agycomment}[1]{\todo[size=\scriptsize, linecolor=orange, bordercolor=orange, backgroundcolor=white]{\textcolor{gfored}{\textbf{@gy:} #1}}}
    \newcommand{\agyinline}[1]{\textcolor{gfored}{\textbf{[@agy: }#1\textbf{]}}}

    \newcommand{\atbcomment}[1]{\todo[size=\scriptsize, linecolor=orange, bordercolor=orange, backgroundcolor=white]{\textcolor{ao}{\textbf{@atb:} #1}}}

    \newcommand{\hluoinline}[1]{\textcolor{moegi}{\textbf{[@hluo: }#1\textbf{]}}}
    \newcommand{\hluobox}[1]{\todo[size=\scriptsize, linecolor=orange, bordercolor=orange, backgroundcolor=white]{\textcolor{moegi}{\textbf{@hluo:} #1}}}

    \newcommand{\yctcomment}[1]{\todo[size=\scriptsize, linecolor=orange, bordercolor=orange, backgroundcolor=white]{\textcolor{yt}{\textbf{@yct:} #1}}}

    \newcommand{\joel}[1]{\textcolor{jl}{#1}}
    \newcommand{\joelcomment}[1]{\todo[size=\scriptsize,linecolor=orange,bordercolor=orange,backgroundcolor=white]{\textcolor{jl}{\textbf{@joel:} #1}}}

\else

    \newcommand{\ominline}[1]{}
    \newcommand{\ombox}[1]{}

    \newcommand{\agycomment}[1]{}
    \newcommand{\agyinline}[1]{}

    \newcommand{\atbcomment}[1]{}

    \newcommand{\hluoinline}[1]{}
    \newcommand{\hluobox}[1]{}

    \newcommand{\yctcomment}[1]{}

    \newcommand{\joel}[1]{{#1}}
    \newcommand{\joelcomment}[1]{}

\fi

\newif\ifrebuttal
\rebuttaltrue

\ifrebuttal
\usepackage[colorinlistoftodos,prependcaption,textsize=small]{todonotes}

\definecolor{darkred}{rgb}{0.9, 0.0, 0.0}

\definecolor{darkblue}{rgb}{0.0, 0.0, 0.85}

\fi

\newcommand*\DRAMCMD[1]{\texttt{#1}}
\newcommand*\DRAMTIMING[1]{t\textsubscript{#1}}

\newcounter{obs}
\newcounter{tkw}
\usepackage{glossaries}

\newacronym{vdd}{$V_{DD}$}{supply voltage}
\newacronym{vpp}{$V_{PP}$}{wordline voltage}
\newacronym{vwl}{$V_{PP}$}{wordline voltage}
\newacronym{vgs}{$V_{GS}$}{gate-to-source voltage}
\newacronym{vth}{$V_{TH}$}{the voltage threshold that the bitline voltage should exceed for the activation to be reliably completed}
\newacronym{gnd}{$GND$}{ground}

\newacronym{ber}{$BER$}{the fraction of DRAM cells that experience bitflips in a DRAM row}
\newacronym{acmin}{$AC_{min}$}{the minimum number of total aggressor row activations to cause at least one bitflip}
\newacronym{ac}{$AC$}{activation count}
\newacronym{rblast}{$r_{Blast}$}{blast radius}
\newacronym{iqr}{$IQR$}{interquartile range}

\newacronym{trcd}{\DRAMTIMING{RCD}}{
{the minimum time between opening a row with an \DRAMCMD{ACT} command and accessing the row buffer}
}
\newacronym{trp}{\DRAMTIMING{RP}}{
{the minimum time between sending a \DRAMCMD{PRE} command and opening a row with an \DRAMCMD{ACT} command}
}
\newacronym{tras}{\DRAMTIMING{RAS}}{
{the minimum time between opening a row with an \DRAMCMD{ACT} command and closing the row with a \DRAMCMD{PRE} command}
}
\newacronym{trefi}{\DRAMTIMING{REFI}}{the \joel{default} time interval \joel{between consecutive \DRAMCMD{REF} commands}}
\newacronym{trefw}{\DRAMTIMING{REFW}}{the maximum time window between two consecutive refresh operations targeting {the same} row}

\def\BibTeX{{\rm B\kern-.05em{\sc i\kern-.025em b}\kern-.08em
    T\kern-.1667em\lower.7ex\hbox{E}\kern-.125emX}}

\newcommand{\affilETH}[0]{\textsuperscript{\S}}
\newcommand{\affilGoogle}[0]{\textsuperscript{$\dagger$}}
\newcommand{\affilOracle}[0]{\textsuperscript{$\ddagger$}}
\newcommand{\affilSNU}[0]{\textsuperscript{$\nabla$}}

\author{%
{Junwhan Ahn}\affilGoogle \hspace{0.5cm} {Sungpack Hong}\affilOracle \hspace{0.5cm} {Sungjoo Yoo}\affilSNU \hspace{0.5cm} {Onur Mutlu}\affilETH \hspace{0.5cm} {Kiyoung Choi}\affilSNU%
\vspace{-3pt}
\\
\affilGoogle\emph{Google DeepMind} 
\hspace{1cm} 
\affilOracle\emph{Oracle Labs} 
\hspace{1cm} 
\affilETH\emph{ETH Z{\"u}rich}%
\hspace{1cm} 
\affilSNU\emph{Seoul National University}%
\vspace{-20pt}
}

\title{\LARGE{\emph{Retrospective:} A Scalable Processing-in-Memory Accelerator \\\vspace{-3pt} for Parallel Graph Processing}\vspace{-10pt}}

\begin{document}
\maketitle
\thispagestyle{plain}
\pagestyle{plain}
\setstretch{0.80}
\begin{abstract}  
Our ISCA 2015 paper~\cite{ahn2015scalable} provides a new programmable processing-in-memory (PIM) architecture and system design that can accelerate key data-intensive applications, with a focus on graph processing workloads. Our major idea was to completely rethink the system, including the programming model, data partitioning mechanisms, system support, instruction set architecture, along with near-memory execution units and their communication architecture, such that an important workload can be accelerated at a maximum level using a distributed system of well-connected near-memory accelerators. We built our accelerator system, Tesseract, using 3D-stacked memories with logic layers, where each logic layer contains general-purpose processing cores and cores communicate with each other using a message-passing programming model. Cores could be specialized for graph processing (or any other application to be accelerated). 

To our knowledge, our paper was the first to completely design a near-memory accelerator system from scratch such that it is both generally programmable and specifically customizable to accelerate important applications, with a case study on major graph processing workloads. Ensuing work in academia and industry showed that similar approaches to system design can greatly benefit both graph processing workloads and other applications, such as machine learning, for which ideas from Tesseract seem to have been influential.

This short retrospective provides a brief analysis of our ISCA 2015 paper and its impact. We briefly describe the major ideas and contributions of the work, discuss later works that built on it or were influenced by it, and make some educated guesses on what the future may bring on PIM and accelerator systems. 

\end{abstract}

\section{Background, Approach \& Mindset}

We started our research when 3D-stacked memories (e.g.,~\cite{hmc.spec.1.1, jeddeloh2012hybrid, jedec.hbm.spec}) were viable and seemed to have promise for building effective and practical processing-near-memory systems. Such near-memory systems could lead to improvements, but there was little to no research that examined how an accelerator could be completely (re-)designed using such near-memory technology, from its hardware architecture to its programming model and software system, and what the performance and energy benefits could be of such a re-design. We set out to answer these questions in our ISCA 2015 paper~\cite{ahn2015scalable}.

We followed several major principles to design our accelerator from the ground up. We believe these principles are still important: a major contribution and influence of our work was in putting all of these together in a cohesive full-system design and demonstrating the large performance and energy benefits that can be obtained from such a design. We see a similar approach in many modern large-scale accelerator systems in machine learning today (e.g.,~\cite{tpu-system, brainwave-system, cerebras-ieeemicro23, tesla-dojo-ieeemicro23, nvidia-superpod}). Our principles are:

1. {\em Near-memory execution} to enable/exploit the high data access bandwidth modern workloads (e.g., graph processing) need and to reduce data movement and access latency.

2. {\em General programmability} so that the system can be easily adopted, extended, and customized for many workloads.

3. {\em Maximal acceleration capability} to maximize the performance and energy benefits. We set ourselves free from backward compatibility and cost constraints. We aimed to completely re-design the system stack. Our goal was to explore the maximal performance and energy efficiency benefits we can gain from a near-memory accelerator if we had complete freedom to change things as much as we needed. We contrast this approach to the {\em minimal intrusion} approach we also explored in a separate ISCA 2015 paper~\cite{pei}.

4. {\em Customizable to specific workloads}, such that we can maximize acceleration benefits. Our focus workload was graph analytics\slash processing, a key workload at the time and today. However, our design principles are not limited to graph processing and the system we built is customizable to other workloads as well, e.g., machine learning, genome analysis.

5. {\em Memory-capacity-proportional performance}, i.e.,  processing capability should proportionally grow (i.e., scale) as memory capacity increases and vice versa. This enables scaling of data-intensive workloads that need both memory and compute. 

6. {\em Exploit new technology (3D stacking)} that enables tight integration of memory and logic and helps multiple above principles (e.g., enables customizable near-memory acceleration capability in the logic layer of a 3D-stacked memory chip). 

7. {\em Good communication and scaling capability} to support scalability to large dataset sizes and to enable memory-capacity-proportional performance. To this end, we provided scalable communication mechanisms between execution cores and carefully interconnected small accelerator chips to form a large distributed system of accelerator chips. 

8. {\em Maximal and efficient use of memory bandwidth} to supply the high-bandwidth data access that modern workloads need. To this end, we introduced new, specialized mechanisms for prefetching and a programming model that helps leverage application semantics for hardware optimization.

\section{Contributions and Influence}

We believe the major contributions of our work were 1) complete rethinking of how an accelerator system should be designed to enable maximal acceleration capability, and 2) the design and analysis of such an accelerator with this mindset and using the aforementioned principles to demonstrate its effectiveness in an important class of workloads. 

One can find examples of our approach in modern large-scale machine learning (ML) accelerators, which are perhaps the most successful incarnation of scalable near-memory execution architectures. ML infrastructure today (e.g.,~\cite{tpu-system, brainwave-system, cerebras-ieeemicro23, tesla-dojo-ieeemicro23, nvidia-superpod}) consists of accelerator chips, each containing compute units and high-bandwidth memory tightly packaged together, and features scale-up capability enabled by connecting thousands of such chips with high-bandwidth interconnection links. The system-wide rethinking that was done to enable such accelerators and many of the principles used in such accelerators resemble our ISCA 2015 paper's approach. 

The ``memory-capacity-proportional performance'' principle we explored in the paper shares similarities with how ML workloads are scaled up today. Similar to how we carefully sharded graphs across our accelerator chips to greatly improve effective memory bandwidth in our paper, today’s ML workloads are sharded across a large number of accelerators by leveraging data\slash model parallelism and optimizing the placement to balance communication overheads and compute scalability~\cite{pope2022efficiently,gshard}. With the advent of large generative models requiring high memory bandwidth for fast training and inference, the scaling behavior where capacity and bandwidth are scaled together has become an essential architectural property to support modern data-intensive workloads.

The ``maximal acceleration capability'' principle we used in Tesseract provides much larger performance and energy improvements and better customization than the “minimalist” approach that our other ISCA 2015 paper on {\em PIM-Enabled Instructions}~\cite{pei} explored: “minimally change” an existing system to incorporate (near-memory) acceleration capability to ease programming and keep costs low. So far, the industry has more widely adopted the maximal approach to overcome the pressing scaling bottlenecks of major workloads. The key enabler that bridges the programmability gap between the maximal approach favoring large performance \& energy benefits and the minimal approach favoring ease of programming is compilation techniques.  These techniques lower well-defined high-level constructs into lower-level primitives~\cite{gshard,collective-matmul}; our ISCA 2015 papers~\cite{ahn2015scalable, pei} and a follow-up work~\cite{aim} explore them lightly. We believe that a good programming model that enables large benefits coupled with support for it across the entire system stack (including compilers \& hardware) will continue to be important for effective near-memory system and accelerator designs~\cite{aim}. We also believe that the maximal versus minimal approaches that are initially explored in our two ISCA 2015 papers is a useful way of exploring emerging technologies (e.g., near-memory accelerators) to better understand the tradeoffs of system designs that exploit such technologies. 

\section{Influence on Later Works}

Our paper was at the beginning of a proliferation of scalable near-memory processing systems designed to accelerate key applications (see~\cite{mutlu2020modern} for many works on the topic). Tesseract has inspired many near-memory system ideas (e.g.,~\cite{graphp,graphr,graphq,dai2018graphh,graphia,rheindt2019nemesys,belayneh2020graphvine,challapalle2020gaas,zhou2021ultra,xie2021spacea, zhou2021hygraph,lenjani2022gearbox,dalorex}) and served as the de facto comparison point for such systems, including near-memory graph processing accelerators that built on Tesseract and improved various aspects of Tesseract. Since machine learning accelerators that use high-bandwidth memory (e.g.,~\cite{tpu-system, nvidia-hopper}) and industrial PIM prototypes (e.g.,~\cite{devaux2019, gomez-luna_benchmarking_2022, upmem-ml, samsung-pim, samsung-pim2, axdimm, axdimm-database, hynix-pim, alibaba-pim, sk-hynix-aim2, singh_fpga-based_2021, singh_accelerating_2021}) are now in the market, near-memory processing is no longer an ``eccentric'' architecture it used to be when  Tesseract was originally published.

Graph processing \& analytics workloads remain as an important and growing class of applications in various forms, ranging from large-scale industrial graph analysis engines (e.g.,~\cite{pgx}) to graph neural networks~\cite{gcn}. Our focus on large-scale graph processing in our ISCA 2015 paper increased attention to this domain in the computer architecture community, resulting in subsequent research on efficient hardware architectures for graph processing (e.g.,~\cite{nai2017graphpim, besta2021sisa, graphicionado}).

\section{Summary and Future Outlook}

We believe that our ISCA 2015 paper's principled rethinking of system design to accelerate an important class of data-intensive workloads provided significant value and enabled\slash influenced a large body of follow-on works and ideas. We expect that such rethinking of system design for key workloads, especially with a focus on ``maximal acceleration capability,'' will continue to be critical as pressing technology and application scaling challenges increasingly require us to think differently to substantially improve performance and energy (as well as other metrics). We believe the principles exploited in Tesseract are fundamental and they will remain useful and likely become even more important as systems become more constrained due to the continuously-increasing memory access and computation demands of future workloads.  We also project that as hardware substrates for near-memory acceleration (e.g., 3D stacking, in-DRAM computation, NVM-based PIM, processing using memory~\cite{mutlu2020modern}) evolve and mature, systems will take advantage of them even more, likely using principles similar to those used in the design of Tesseract.

\setstretch{0.75}

\balance
\bibliographystyle{IEEEtran}
{\tiny
\bibliography{combined}}

\end{document}